# Suppressing the Errors due to Mode Mismatch for *M*-ary PSK Quantum Receivers by Photon-Number-Resolving Detector

Ke Li, Yuan Zuo, and Bing Zhu

***Abstract*—A *M*-ary phase shift keying (PSK) quantum receiver consisting of displacement operations, photon counting, and electrical feed-back adaptive measurements is analyzed with a realistic model considering the effects of the sub-unity quantum efficiency and the dark counts of single-photon detectors, as well as the transmittance and the mode mismatch of beam splitters. Among these factors, the mode mismatch has the greatest impact on the error probability of the receiver with on-off detectors. The errors due to mode mismatch can be suppressed effectively by using photon-number-resolving detectors (PNRD) instead of on-off detectors.**

## I. Introduction

QUANTUM receiver can discriminate different coherent states with an error probability smaller than the conventional limit (standard quantum limit: SQL) which is attained by classical receiver (direct detection, homodyne and heterodyne receiver). It plays an important role for coherent states quantum key distribution (QKD) and future optical communications approaching the ultimate limit allowed by the laws of quantum mechanics.

Since Helstrom obtained the quantum mechanical bound of the minimum error probability, which is significantly lower than the SQL [1], physical implementation of the quantum receiver to achieve the Helstrom bound have been extensively investigated for different coherent state modulation sets. For binary modulation, Kennedy receiver [2], Dolinar receiver [3], [4] and Optimal Displacement Receiver [5] have been theoretically proposed and experimentally demonstrated successively. Recently, Partitioned-Interval Detection binary quantum receiver [6], [7] was theoretically analyzed to balance the operation rate and the receiver performance. For multiple modulation, quantum receivers respectively based on classic-quantum hybrid structure [8] and adaptive measurements [9], [10] have also been investigated for *M*-ary phase shift keying (PSK) [9] and pulse position modulation (PPM) signals [10]. All of the above mentioned receivers contain two key parts, and they are displacement operation and photon counting, which can be implemented by beam splitters and single-photon detectors (on-off detectors or photon-number-resolving detectors: PNRD). In practice, the sub-unity quantum efficiency and the dark counts of single-photon detectors, as well as the transmittance and the mode mismatch of beam splitters will deteriorate the performances. Recently, the robustness of the PNRD based quantum receiver against the dark counts was proved through Monte Carlo simulation [11]. However, previous experiments showed that the dark counts had little effect on the performance of the receiver, while the mode mismatch was the major affecting factors [9].

In this letter, we will theoretically study each of the above four non-ideal factors' impacts on the error probability of the *M*-ary PSK adaptive measurements feed-back quantum receiver. And it will be proved by Monte Carlo simulation that, besides the robustness against the dark counts, with PNRD instead of on-off detectors, the errors due to mode mismatch can also be suppressed effectively, especially for higher signal mean photon numbers.

## II. Receiver Modeling

In this section, let us consider the modeling of *M*-ary PSK adaptive measurements feed-back quantum receiver. Here we emphasize the PNRD based receiver modeling including the mode mismatch effects, whose performance is not clear so far. The *M*-ary PSK coherent states signals to be discriminated are defined as

$$\left|\alpha_m\right\rangle = \left|\alpha u^m\right\rangle, \quad u = e^{2\pi i/M}, \tag{1}$$

where $M = 0,1,\ldots,M-1$. Without loss of generality, $\alpha$ is chosen to be real, and the mean photon numbers of the signals are $|\alpha|^2$. Throughout this paper, *a priori* probabilities of these coherent states are assumed to be equal for all $m$, i.e. $P_m = 1/M$.

Fig. 1 shows the schematic of the adaptive measurements feed-back quantum receiver. Firstly, each received signal

This Manuscript received ×××; revised ×××; accepted××××. Date of publication ×××; date of current version ×××. This work was supported by ×××.

The authors are with the Department of Electronic Engineering and Information Science, University of Science and Technology of China, Hefei, Anhui 230027, China (e-mail: zbing@ustc.edu.cn).

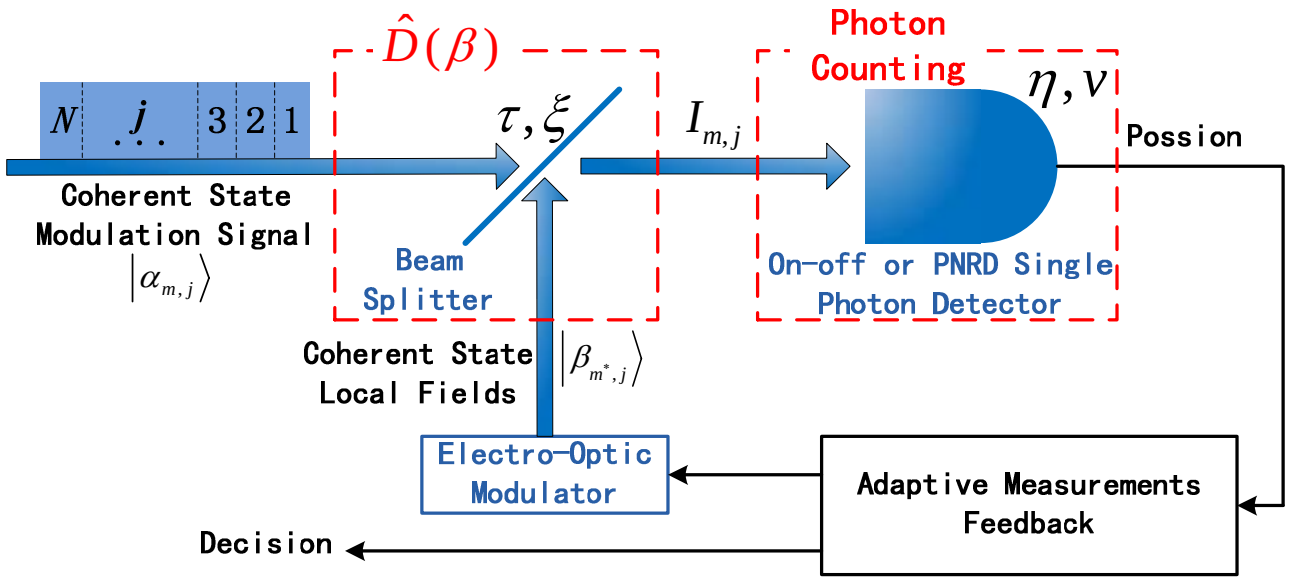

Fig. 1. The schematic of the adaptive measurements feed-back quantum receiver. $\xi$ describes the mode mismatch and $\tau$ is the transmittance of the beam splitter. $\nu$ describes the dark counts and $\eta$ is the quantum efficiency of the single-photon detector. The single-photon detector can be on-off detector or PNRD. And the local field is updated according to the maximum *a posteriori* criterion. In the figure, the thick blue arrows indicate the optical signal and the thin black arrows indicate the electrical signal.

codeword interval is partitioned into $N$ consecutive disjoint equal durations with indexes $j=1,2,3,\ldots,N$, and the signals corresponding to each duration are

$$\left|\alpha_{m,j}\right\rangle=\left|\frac{\alpha_m}{\sqrt{N}}\right\rangle. \tag{2}$$

Secondly, the signal of the each duration is displaced by a local field $\left|\beta_{m^*,j}\right\rangle$ via a beam splitter, and the average intensity $I_{m,j}$ of the field after the beam splitter in the photon number units is

$$I_{m,j}=(1-\xi)(\tau\left|\alpha_{m,j}\right|^2+\left|\beta_{m^*,j}\right|^2)+\xi\left|\sqrt{\tau}\alpha_{m,j}-\beta_{m^*,j}\right|^2, \tag{3}$$

where $\xi$ describes the mode mismatch and $\tau$ is the transmittance of the beam splitter. If $j=1$, the receiver tries to null the received signal with the local field $\beta_{m^*,j}=\sqrt{\tau}\alpha_{0,1}=\sqrt{\tau}\alpha_0/\sqrt{N}$, which means the receiver starts by hypothesis $\left|\alpha_0\right\rangle$ in the first duration of each codeword. If $j>1$, the local field $\beta_{m^*,j}=\sqrt{\tau}\alpha_{m^*,j-1}$, and $m^*$ is decided by the adaptive measurement strategy to be introduced in the following.

Next, the number of photons $n_j$ of the displaced fields is detected by using a single-photon detector. If the single-photon detector is a PNRD, the probability of detecting $n_j$ photons for the displaced field in each duration is given by

$$P_{m,j}=e^{-\nu-\eta I_{m,j}}\frac{(\nu+\eta I_{m,j})^{n_j}}{n_j!}, \tag{4}$$

where $\nu$ describes the dark counts and $\eta$ is the quantum efficiency of the single-photon detector. Finally, the local field $\left|\beta_{m^*,j+1}\right\rangle$ is updated according to the maximum *a posteriori* (MAP) in the $j$th duration. The *a posteriori* probability for after detecting $n_j$ photons in the $j$th duration is given by

$$P_{post-m,j}=\frac{P_{prior-m,j}\cdot P_{m,j}}{\sum_{l=0}^{M-1}P_{prior-l,j}\cdot P_{l,j}}, \tag{5}$$

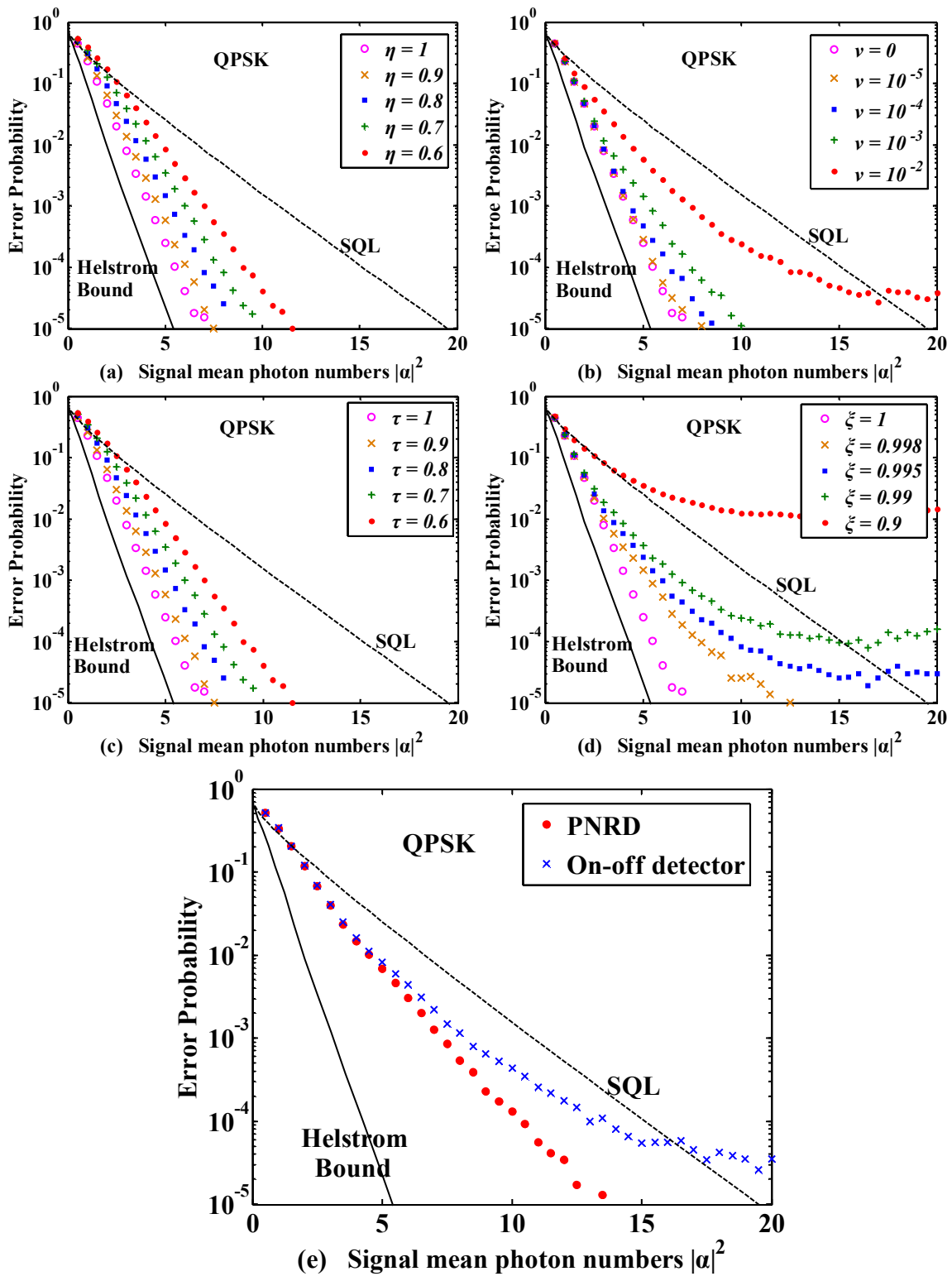

Fig. 2. The separate effects of (a) the quantum efficiency ($\eta$ varies with $\nu=0$, $\tau=1$, and $\xi=1$) and (b) the dark counts ($\nu$ varies with $\eta=1$, $\tau=1$, and $\xi=1$) of the single-photon detector, as well as (c) the transmittance ($\tau$ varies with $\eta=1$, $\nu=0$, and $\xi=1$) and (d) the mode mismatch ($\xi$ varies with $\eta=1$, $\nu=0$, and $\tau=1$) of the beam splitter for QPSK quantum receiver with on-off detector. (e) shows the QPSK quantum receiver performance with $\eta=0.723$, $\nu=2.7\times10^{-5}$, $\tau=0.99$, and $\xi=0.995$ using different detectors. In both figures, the black solid and dashed lines represent the Helstrom bound and the standard quantum limits, respectively.

The *a priori* probability $P_{prior-m,j}$ in the $j$th duration satisfies

$$\begin{cases}P_{prior-m,j}=P_m, & j=1\\ P_{prior-m,j}=P_{post-m,j-1}, & j>1\end{cases} \tag{6}$$

Then $m^*$ of the local field $\left|\beta_{m^*,j+1}\right\rangle$ can be obtained by

$$P_{post-m^*,j}=\max_m(P_{post-m,j}). \tag{7}$$

And the $m^*$ in the last duration corresponding to $P_{post-m^*,N}$ is the decision output.

If the single-photon detector is an on-off detector observing only zero or nonzero photons, the detector can be described by the following probabilities

$$\begin{cases}P_{off-m,j}=e^{-\nu-\eta I_{m,j}}\\ P_{on-m,j}=1-e^{-\nu-\eta I_{m,j}}\end{cases}, \tag{8}$$

$P_{off-m,j}$ is the probability of finding an off signal and $P_{on-m,j}$ is the probability of finding an on signal in each duration. In this situation, when we calculate the *a posteriori* probability, $P_{m,j}$ in equation (5) becomes the following expressions

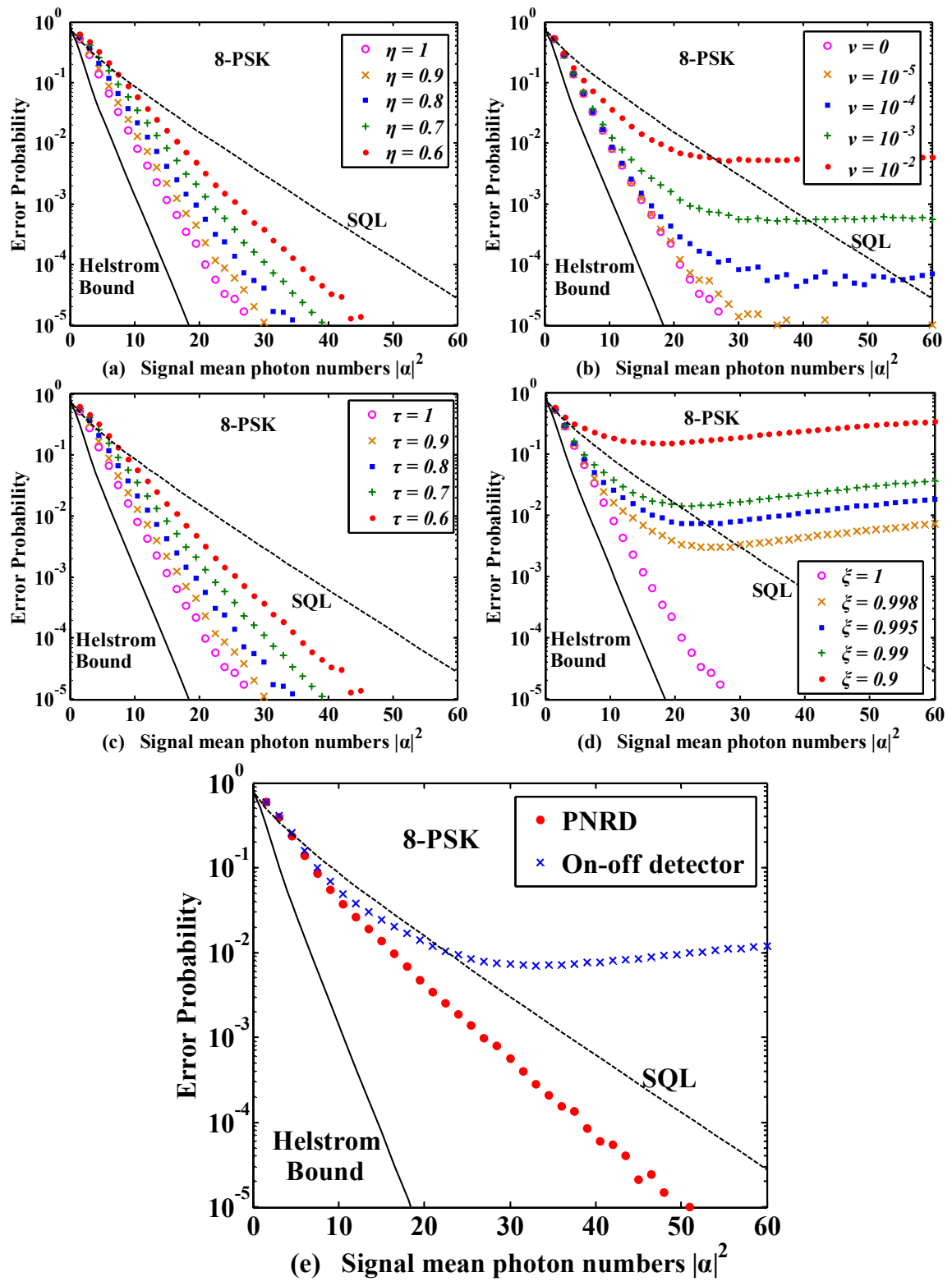

Fig. 3. The separate effects of (a) the quantum efficiency ( $\eta$ varies with $v=0$ , $\tau=1$, and $\xi=1$ ) and (b) the dark counts ( $v$ varies with $\eta=1$, $\tau=1$, and $\xi=1$ ) of the single-photon detector, as well as (c) the transmittance ( $\tau$ varies with $\eta=1$, $v=0$ , and $\xi=1$ ) and (d) the mode mismatch ( $\xi$ varies with $\eta=1$, $v=0$ , and $\tau=1$) of the beam splitter for 8-PSK quantum receiver with on-off detector. (e) shows the 8-PSK quantum receiver performance with $\eta=0.723$ , $v=2.7\times10^{-5}$ , $\tau=0.99$ , and $\xi=0.995$ using different detectors. In both figures, the black solid and dashed lines represent the Helstrom bound and the standard quantum limits, respectively.

$$\begin{cases} P_{m,j} = e^{-v-\eta I_{m,j}}, & n_j = 0, \quad off \\ P_{m,j} = 1 - e^{-v-\eta I_{m,j}}, & n_j > 0, \quad on \end{cases} \tag{9}$$

other than equation (4).

After establishing this probability model, we can study the performance of the receiver by using the Monte Carlo simulation.

## III. Simulation Results

In this section, let us discuss the receiver performance simulation results for QPSK and 8-PSk signals. For all simulations, the number of partitions is $N=10$. Each plot is given by a Monte Carlo simulation with $10^6$ trials.

Fig. 2 illustrates the simulation results for QPSK signals. Fig. 2 (a), (b), (c) and (d) shows the separate effects of the four imperfect factors on the receiver error probability. Form Fig. 2 (a) and (c), it is clear that the quantum efficiency of the on-off detector and the transmittance of the beam splitter have the same effect on the receiver performance. From Fig. 2 (b), we note that the dark counts of the on-off detector will saturate the error probability as the signal mean photon numbers increase. And when the dark count rate is small, the effect is not obvious. From Fig. 2 (d), it is shown that the receiver performance deteriorates seriously even with a minor mode mismatch. So in practice, we should give more attention to the mode mismatch between the signal and the local fields. Fig. 2 (e) shows the overall effects of the imperfect factors on the receiver performance with on-off detector and PNRD, and the imperfect factors in the simulation correspond to the experimental conditions of reference [9]. From the above analysis, it is clear that the upward curvature seen in the simulation with on-off detector in Fig. 2 (e) is mainly due to the mode mismatch. As illustrated in Fig. 2 (e), the errors due to mode mismatch are suppressed effectively by using PNRD instead of on-off detector, especially for larger signal mean photon numbers.

Fig. 3 shows the simulation results for 8-PSK signals. The curves in Fig. 3 are similar to the ones in Fig. 2, and the conclusions for both Fig. 2 and Fig. 3 are the same.

## IV. Discussions

In this letter, it is shown that the errors of *M*-ary PSK quantum receiver due to mode mismatch can be suppressed effectively by using PNRD. But it should be mentioned that the PNRD modeled in the article is ideal, and the finite resolution and PNR capability of the PNRD are not considered.